\documentclass[aps,prd,twocolumn,groupedaddress,showpacs]{revtex4}
\usepackage{graphicx}
\usepackage{dcolumn}
\usepackage{bm}
\usepackage{amsmath}
\usepackage{epstopdf}
\usepackage{amsfonts}
\usepackage{amssymb}%

\usepackage[colorlinks=true,citecolor=blue,urlcolor=magenta,breaklinks]{hyperref}

\usepackage{natbib}

\providecommand{\U}[1]{\protect\rule{.1in}{.1in}}

\begin{document}

\title{Constraining the scalar singlet and inert dark matter \\
models using neutron stars}

\author{Grigorios Panotopoulos and Il\'\i dio Lopes}
\email{grigorios.panotopoulos@tecnico.ulisboa.pt, ilidio.lopes@tecnico.ulisboa.pt}
\affiliation{CENTRA, Instituto Superior T{\'e}cnico,\\ Universidade de Lisboa,
Av. Rovisco Pa{\'i}s 1, Lisboa, Portugal}

\date{\today}

\begin{abstract}
In the present work we study the scalar singlet as well as the Two-Higgs Doublet model inert dark matter particles impact on compact objects, and we provide the first constraints of the parameter space using neutron stars. The models discussed here are
characterized by two free parameters, namely the mass $M_\chi$ of the scalar particle that plays the role of the dark matter in the Universe, and a dimensionless coupling constant $\lambda_\chi$ that determines the strength of the interaction of the dark matter particles with the Standard Model Higgs boson. By considering a typical neutron star we were able to obtain constraints on scalar dark matter depending on the DM annihilation cross section and self-interaction coupling constant. Our findings show that i) for heavy DM particles neutron stars can provide us with bounds better that the current limits from direct detection searches only when the self-annihilations of DM particles are negligible and the DM self-interaction coupling constant is very small, while ii) for light DM particles the bounds obtained here are comparable to limits from Higgs invisible decays unless the DM particles are extremely light.
\end{abstract}

\maketitle

\section{Introduction}

Several well-established observational and theoretical results from Cosmology and Astrophysics strongly suggest that the non-relativistic matter component of the Universe is dominated by a new type of matter particles, yet to be discovered, the so-called Dark Matter (DM). It was in 1933 when Zwicky studying clusters of galaxies introduced the term "missing mass" or dark matter \cite{zwicky}. Much later Rubin and Ford with optical studies of M31 made the case for DM in galaxies in 1970 \cite{rubin}. For a review on dark matter see e.g. \cite{munoz}. Despite the fact that as of today there are many DM candidates \cite{taoso}, the nature and origin of DM still remains a mystery, comprising one of the biggest challenges in modern theoretical cosmology. Among all the possible choices,
perhaps the most popular class of DM candidates is the so-called Weakly Interacting Massive Particles (WIMPs), that are thermal relics from the Big-Bang. Initially the temperature of the Universe was high enough to maintain the DM particle in equilibrium with the rest of the particles. However, as the Universe expands and cools down at some point the annihilation rate of DM particles
$\Gamma=n_\chi \langle \sigma v \rangle_\chi$,  with $n_\chi$ being the number density of the DM particle $\chi$
and $\langle \sigma v \rangle_\chi$ the thermal average annihilation cross section, drops below the Hubble parameter $H$
that measures the expansion rate of the Universe. When this happens the DM abundance freezes-out since
the $\chi$ particles can no longer annihilate, and their current abundance remains the same ever since.
It turns out that their today's relic density is given by \cite{SUSYDM}
\begin{equation}
\Omega_{\chi} h^2 = \frac{3 \times 10^{-27} cm^3/s}{\langle \sigma v \rangle_\chi}
\end{equation}
where $h$ is related to the Hubble constant $H_0=100 \: h (km s^{-1})/(Mpc)$. In order to reproduce the observed DM abundance $\Omega_\chi h^2=0.1198 \pm 0.0015$ \cite{wmap,planck2015}, the WIMPs annihilation cross section must have the value $\langle \sigma v \rangle_{st} \simeq 3 \times 10^{-26} cm^3/s$, which is a typical value for a particle that does not have neither strong nor electromagnetic interactions. One should keep in mind that this result is obtained assuming a cosmological scenario with a high reheating temperature $T_R$ after inflation, in which the DM particle is a thermal relic from the Big-Bang. However, the reheating temperature does not have to be high, as primordial Big-Bang nucleosynthesis and thermalization of all 3 neutrino species requires $T_R > 4 MeV$ \cite{hannestad}, and in fact in the literature various cosmological scenarios with a low reheating temperature have been studied \cite{lowreh1,lowreh2,lowreh3,lowreh4}. In these scenarios non-thermal production mechanisms for the DM particles are invoked, and thus the DM abundance can be reproduced even if the DM particle annihilation cross section does not have the "standard" value.

In this work we will focus our study in two special classes of WIMPs known as scalar (inert or singlet) dark matter particles,
which are the simplest and most economical extensions of the Standard Model (SM) of particle physics.
In the first class \cite{singlet1,singlet2,singlet3,singlet4} the scalar sector consists of the SM Higgs boson as well as a real scalar $S$ that
is gauge singlet and it does not have direct interactions with fermions. Furthermore, the extra scalar field is stable due to a discrete $Z_2$ symmetry, and since it is neutral it is a very good DM candidate. In the second class \cite{inert1,inert2,inert3} the Higgs sector consists of two Higgs doublets $H_1,H_2$, while a discrete $Z_2$ symmetry forbids the Yukawa couplings for the second doublet.
After electroweak symmetry breaking 5 physical scalar fields remain in the spectrum, namely the SM Higgs $h$, two charged bosons $H^{\pm}$, a CP even $H^0$ and
a CP odd $A^0$ scalars that are neutral. The CP even scalar, if it the lightest among the extra
Higgs bosons, becomes stable and therefore it can play the role of the DM in the Universe.

To probe the nature of dark matter several earth based experiments have been designed.
In these experiments an effort is made to observe the nucleus recoil after a dark matter particle scatters
off the material of the detector. These direct detection experiments have put limits on the DM-nucleon
candidate cross section for a given mass of the DM particle \cite{detection1,detection1b,detection2}.
During the last 15 years or so observational data from astrophysical objects, such as the Sun~\cite{ilidio1,ilidio2,ilidio3}, solar-like stars~\cite{ilidio4,ilidio5,ilidio6}, white dwarfs and neutron stars~\cite{kouvaris0, kouvaris1, kouvaris2}, have
been employed to offer us complementary bounds on the DM-nucleon cross section, see e.g.~\cite{ilidio0} and references
therein.

\begin{figure}[ht!]
\centering
\includegraphics[scale=1]{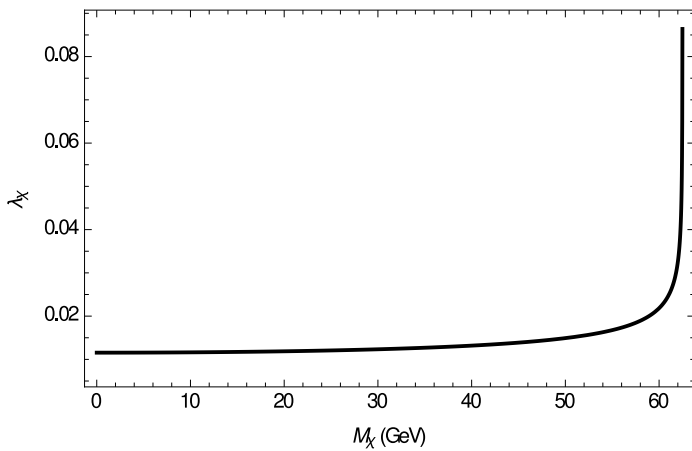}
\caption{Shown is the constraint from invisible SM Higgs decays $BR(h \rightarrow inv.) \leq 0.3$ \cite{invisible} in the $M_\chi$ -- $\lambda_\chi$ plane.
The allowed parameter space lies below the curve.}
\label{fig:1} 	
\end{figure}

Since neutron stars are the densest objects in the universe after black holes, they comprise excellent laboratories to study and constrain non-standard physics. It is the aim of the present article to use neutron stars to constrain the parameter space of the scalar dark matter. Our work is organized as follows: after this introduction, we present the theoretical framework
in section two, and we constrain the scalar parameter space in the third section. Finally we conclude in section four.
We work in units in which the speed
of light in vacuum $c$, the Boltzmann constant $k_B$ and the reduced Planck mass $\hbar$ are set equal to unity.
In these units all dimensionful quantities are measured in GeV, and we make use of the conversion rules
$1 m = 5.068 \times 10^{15} GeV^{-1}$, $1 kg = 5.610 \times 10^{26} GeV$ and $1 K = 8.617 \times 10^{-14} GeV$ \cite{guth}.

\section{Theoretical framework}


\begin{figure}[ht!]
	\centering
	\includegraphics[scale=1]{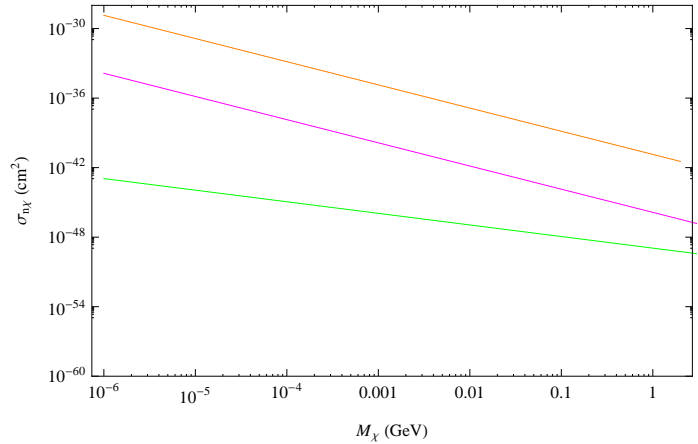}
	\caption{DM-nucleon cross section in $cm^2$ versus scalar mass $M_\chi$ in GeV for case I, light scalar DM particles and for the 3 values of self-interaction coupling (green curve for $\lambda_1=0$, magenta curve for $\lambda_2=10^{-30}$, and orange curve for $\lambda_3=10^{-21}$). The solid curves correspond to the formation of a mini black hole, while in this mass range there are no limits from direct detection experiments. To avoid the formation of a black hole inside the star, for a given $M_\chi$ the cross section must lie below the curves.}
\label{fig:2} 	
\end{figure}

\subsection{The DM-nucleon cross section and Higgs invisible decays}

The DM particles once trapped inside the neutron star interact with the neutrons and eventually thermalize, and since they are
non-relativistic they are described by the Maxwell-Boltzmann distribution~\cite{ilidio1,ilidio1a,ilidio5}. If a large number of them is accreted during the lifetime of a neutron star, they may collapse
and form a mini black hole inside the star that eventually destroy the compact object \cite{goldman}. Therefore, the existence of old neutron stars can impose constraints on the properties of scalar singlet DM. It thus becomes
clear that the most important quantity for the discussion is the DM-nucleon cross section $\sigma_{n\chi}$, which from the theory side can be computed in terms of the free parameters of the model, namely the scalar singlet mass $M_\chi$ and the coupling constant $\lambda_\chi$, while from the experiment side is constrained from direct detection searches, roughly $\sigma_{n\chi}  < 10^{-44} cm^2$ \cite{detection1,detection1b,detection2}.

In the two classes of models analyzed here, the relevant interaction Lagrangian has two terms, namely i) the DM self-interactions
\begin{equation}
\mathcal{L}_{self-int}= -\lambda \chi^4,
\end{equation}
where $\lambda$ is a dimensionless coupling constant,
and ii) the interaction between the SM Higgs boson
$h$ and the DM particle $S$ (singlet scalar DM model) or $H^0$ (inert DM model) \cite{singlet2,inert1}
\begin{equation}
\mathcal{L}_{h \chi \chi}= -\lambda_{\chi} V h \chi^2,
\end{equation}
where $\lambda_{\chi}$ is another dimensionless coupling constant, $V=246 GeV$ is the vacuum expectation value of the SM Higgs boson, and $\chi=S$ or $\chi=H^0$ depending on the model. The relevant Feynman diagram for the process $\chi  N \rightarrow \chi N$, with N being the nucleon, is the one with the SM Higgs exchange. We remark here that in the class of the inert scalar dark matter there is another diagram with the Z boson exchange, but this has already been ruled out from observations \cite{inert1}. Neglecting the difference between neutron and proton, the DM-nucleon scattering cross section
is given by \cite{singlet2,inert1}
\begin{equation}
\sigma_{n\chi} = \frac{\lambda_\chi ^2 f^2 m_n^2 \mu_\chi ^2}{\pi m_h^4 M_\chi ^2}
\end{equation}
and it is spin-independent. In the expression above $m_h=125 GeV$ is the mass of the SM Higgs boson, $M_\chi$ is the DM particle mass, $m_n$ is the nucleon mass, taken to be equal to the mass of the proton  $m_p \simeq 1 GeV$, $\mu_\chi =M_\chi m_n/(M_\chi+m_n)$ is the reduced mass of the DM-nucleon system, and $f$ parameterizes the Higgs-nucleon coupling. A complete expression for the factor $f$ can be found e.g. in \cite{singlet3}. Following the lattice computations \cite{lattice1,lattice2,lattice3} we shall consider the central value $f=0.3$ in agreement with other studies \cite{singlet2,singlet3,singlet4}.

Although neutron stars are hot upon formation, they gradually cool down. However, even isolated neutron stars cannot go below
$10^5 K$ (in agreement with the value taken in \cite{kouvaris2}) due to heating by accretion of interstellar matter \cite{goldman,cooling}. Therefore in the
discussion below we shall consider a typical neutron star (millisecond pulsar) with the following parameters \cite{OV,pulsar1,pulsar2}: mass $M_{\star}  \sim 2 M_{\odot}  = 4 \times 10^{30} kg$, radius $R_{\star}  \sim 10 km$, interior temperature $T_\star \sim 10^5 K$, age $t_* \sim 1 Gyr$, ordinary matter
density $\rho_\star \sim 10^{17} kg/m^3$ and pressure $P_\star \sim 10^{33} N/m^2$, with $M_{\odot}$ being the solar mass.

We remark here that while in the singlet scalar DM model the whole parameter space consists of the three parameters considered here, namely $M_\chi$, $\lambda_\chi$, and
$\lambda$, the parameter space of the inert model involves more parameters. However, in our analysis here only the aforementioned parameters are relevant for the discussion, while the rest of the parameter space is left unconstrained.

Finally, when the decay channel $h \rightarrow \chi \chi$ is kinematically allowed ($m_h > 2 M_\chi$), it contributes to the SM
Higgs boson invisible decays, which by now is constrained from studies at the LHC to be $BR(h \rightarrow inv.) \leq 0.3$ \cite{invisible}. The branching ratio of invisible decays is given by
\begin{equation}
BR(h \rightarrow inv.) = \frac{\Gamma_{inv}}{\Gamma_{SM}+\Gamma_{inv}}
\end{equation}
where $\Gamma_{inv}$ in the classes of models discussed here is given by \cite{singlet4}
\begin{equation}
\Gamma_{inv} = \frac{\lambda_\chi^2 V^2}{8 \pi m_h} \sqrt{1-4 \frac{M_\chi^2}{m_h^2}}
\end{equation}
with $\Gamma_{SM} \simeq 6 MeV$ being the Higgs decay width in the framework of the SM \cite{width}. The constraint $BR(h \rightarrow inv.) \leq 0.3$ in the $M_\chi-\lambda_\chi$ plane is shown in Fig.~\ref{fig:1}.
For a given scalar mass $M_\chi$ the coupling constant $\lambda_\chi$ must lie below the curve.

\subsection{The conditions required for the formation of the black hole}

To see if there is enough DM accretion to collapse and form a black hole inside the star, we need to compute the accretion rate \cite{goldman,kouvaris1}
\begin{equation}
F_\chi =\frac{8 \pi^2 }{3}  \frac{\rho_{\chi} }{M_\chi}  G M_{\star}  R_{\star}
\left( \frac{3}{2 \pi v_\chi^2}\right)^{3/2}  v_\chi^2 \left(1-e^{-3 \frac{E_0}{v_\chi^2}}\right) p
\end{equation}
where we have adopted a DM mean velocity in the neighborhood of the neutron star $v_\chi=270 km/sec$, G is Newton's constant, $E_0=2 (m_p/M_\chi) G M_{\star} /R_{\star} $ is the maximum energy per DM mass that can lead to
capture,  $\rho_{\chi}$ is the local dark matter density (for isolated neutron stars) taken to be $\rho_{\chi}=0.3 GeV/cm^3$. This value for $\rho_{\chi}$ is conservative since current observations suggest
$\rho_{\chi} \simeq 0.38 GeV/cm^3$, while some others indicate a value two times larger (see \cite{ilidio1,ilidio2} for details). Finally the probability p
is given by $p=0.89 \sigma_{n\chi}/\sigma_{cr}$, where the critical cross section is given by
\begin{equation}
\sigma_{cr} = 4 pb \left( \frac{R_{\star} }{R_{\odot} } \right)^2 \left( \frac{M_{\star} }{M_{\odot} } \right)^{-1}=4 \times 10^{-46} cm^2
\end{equation}
while p saturates to unity if $\sigma_{n\chi} > \sigma_{cr}$. Then the accumulated number of DM particles $N_{\chi acc}$ is determined by solving the rate equation \cite{SUSYDM}
\begin{equation}
\frac{dN_{\chi acc}}{dt} = F_\chi - \frac{\langle \sigma v \rangle_\chi}{V_b} \: N_{\chi acc}^2,
\end{equation}
where $V_b$ is the volume of the sphere in which the DM particles are mostly concentrated, and $\langle \sigma v \rangle_\chi$ is the DM
particle annihilation cross section, and it does not necessarily coincide with the classical value required to reproduce the observed DM abundance in eq. (1). With the initial condition $N_{\chi acc}(0)=0$, the rate equation
can be easily integrated, and thus the number of DM particles accumulated inside the star during its lifetime is given by
\begin{equation}
N_{\chi acc} = \sqrt{\frac{F_\chi V_b}{\langle \sigma v \rangle_\chi}}
\: \tanh{\left( \sqrt{\frac{F_\chi \langle \sigma v \rangle_\chi}{V_b}} \: t_*\right)}.
\end{equation}
where $t_*$ is of the order of the Gyr, and gives an estimate of the age of the neutron star \cite{pulsar1,pulsar2}.
It is worth mentioned that the exact solution above acquires a simpler form in two limiting cases, namely when the argument of the function $tanh(x)$ is very small
$x \ll 1$, and also when it is large $x \gg 1$. In the first case one finds
$N_{\chi acc} \simeq F_\chi t_*$, which can be obtained from the rate equation neglecting the annihilation term, while in the second case one finds
\begin{equation}
N_{\chi acc} \simeq  \sqrt{\frac{F_\chi V_b}{\langle \sigma v \rangle_\chi}},
\end{equation}
which can be obtained from the rate equation setting $dN_{\chi acc}/dt=0$. In the following we shall consider these two cases separately, namely first
we shall assume that DM annihilations have a negligible affect (case I), and then we shall consider the case where $\langle \sigma v \rangle_\chi=10^{-33} \langle \sigma v \rangle_{st}$ (case II), where $\langle \sigma v \rangle_{st}$ is the "standard" value required to reproduce the DM abundance assuming a thermal relic from the Big-Bang.

For a gravitational collapse to take place inside the star the following three conditions have to be satisfied:

- First, in a system of non-interacting bosons only the uncertainty principle opposes the collapse. The critical mass of a self-gravitating lump
that can form a black hole is given by \cite{kouvaris2}
\begin{equation}
M_{cr} = \frac{2 M_p^2}{\pi M_\chi} \: \sqrt{1+\frac{\lambda M_p^2}{32 \pi M_\chi^2}}
\end{equation}
with $M_p$ being the Planck mass, and $\lambda$ the DM self-interaction coupling constant. Thus, the first condition to be satisfied is
\begin{equation}
M_{\chi acc} > M_{cr}.
\end{equation}
We should remark here that in the singlet dark matter model there is just one dimensionless coupling constant that determines both the DM self-interactions and the interaction of the DM particle with the SM Higgs boson, while in the inert model the two couplings are independent. In the following we shall consider 3 cases, namely $\lambda_2=10^{-30},\lambda_3=10^{-21}$ and for comparison $\lambda_1=0$, compatible
with observational constraints on self-interacting DM \cite{bulletcluster}.

- The second condition comes from the fact that the newly formed black hole must not emit Hawking radiation \cite{HR1,HR2} too fast. In fact, in the black hole mass rate the Bondi accretion term \cite{bondi} must dominate over the energy loss due to the Hawking radiation \cite{kouvaris2}
\begin{equation}
\frac{4 \pi \rho G^2 M_{\chi acc}^2}{c_s^3} > \frac{1}{15360 \pi G^2 M_{\chi acc}^2}
\end{equation}
with $c_s$ being the speed of sound. Assuming a polytropic equation of state for a non-relativistic Fermi gas $P(\rho)=K \rho^{5/3}$ \cite{OV} the
speed of sound $c_s^2 = dP/d\rho$ is computed to be $c_s=\sqrt{(5 P_\star)/(3 \rho_\star)} \simeq 0.42$. This implies that the second condition is
\begin{equation}
M_{\chi acc} > \left( \frac{c_s^3 M_p^8}{4 \pi^2 \rho_\star \times 15360} \right)^{1/4} = 1.95 \times 10^{37} GeV=M_2
\end{equation}

- Finally, the last condition comes from the onset of DM self-gravitation.
When the total DM mass captured inside a sphere of radius $r_*$ exceeds the mass of the ordinary matter within the same radius
\begin{equation}
M_{\chi acc} > \frac{4 \pi \rho_\star r_*^3}{3}
\end{equation}
the self-gravitation of DM dominate over that of the star \cite{kouvaris2}. Naively it is expected that most of
the DM particles are concentrated inside a radius $r_{th}$ given by \cite{kouvaris1}
\begin{equation}
r_{th}=\left( \frac{9 T_\star}{8 \pi G M_\chi \rho_\star} \right)^{1/2}
\end{equation}
However, as first pointed out by Bose \cite{bose} and later expanded by Einstein \cite{einstein1,einstein2}, in a quantum gas made of bosons
the indistinguishability of the particles requires a new statistical description, now known as Bose-Einstein statistics.
If the temperature of the gas is low enough or the number density of particles is large enough, a new exotic form of matter
is formed. The Bose-Einstein Condensate (BEC) is driven purely by the quantum statistics of the bosons, and not by the interactions between them. The critical temperature is given by \cite{thesis}
\begin{equation}
T_c = \frac{2 \pi \hbar^2}{M_\chi k_B} \left( \frac{n_\chi}{\zeta(3/2)} \right)^{2/3} \simeq 3.3 \frac{n_\chi^{2/3}}{M_\chi}
\end{equation}
in our natural units, where $\zeta(3/2) \simeq 2.612$ is Riemann's zeta function, and $n_\chi=(3 N_{\chi acc})/(4 \pi r_c^3)$ is the number density of the
DM particles. The BEC, considered to be the fifth state of matter after gases, liquids, solids and plasma, is manifested in the classical example of the Helium-4 superfluidity \cite{superfluid}, and led to the Nobel Prize in Physics in 2001 \cite{nobel}.
The size of the condensed state is determined by the radius of the wave function of the scalar singlet ground state in the gravitational potential of the star \cite{kouvaris2}
\begin{equation}
r_c = \left( \frac{8 \pi G \rho_\star M_\chi^2}{3} \right)^{-1/4}
\end{equation}

\section{Constraints on the scalar DM parameter space}

First we employ the thermalization
condition $t_2 < t_*$ derived and used in \cite{kouvaris1}, with $t_2$ given by
\begin{equation}
\frac{t_2 }{4 yr}= \left( \frac{M_\chi}{TeV} \right)^{3/2} \left( \frac{10^8 g/cm^3}{\rho_\star} \right) \left( \frac{10^{-43} cm^2}{\sigma_{n\chi}} \right)
\left( \frac{10^7 K}{T_\star} \right)^{1/2}
\end{equation}
The thermalization condition implies a lower limit for the DM-nucleon cross section
\begin{equation}
\frac{\sigma_{n\chi}}{10^{-52} cm^2} > 4 \left( \frac{M_\chi}{TeV} \right)^{3/2} \left( \frac{10^8 g/cm^3}{\rho_\star} \right) \left( \frac{10^7 K}{T_\star} \right)^{1/2}
\end{equation}
Furthermore, the BEC is formed below the critical
temperature, $T_\star < T_c$, so the condition for its formation is set by
\begin{equation}
\frac{3 N_{\chi acc}}{4 \pi r_c^3} > \left(\frac{M_\chi T_\star}{3.3} \right)^{3/2}
\end{equation}
Our main results are summarized in the figures below.

\begin{figure}[ht!]
	\centering
	\includegraphics[scale=1]{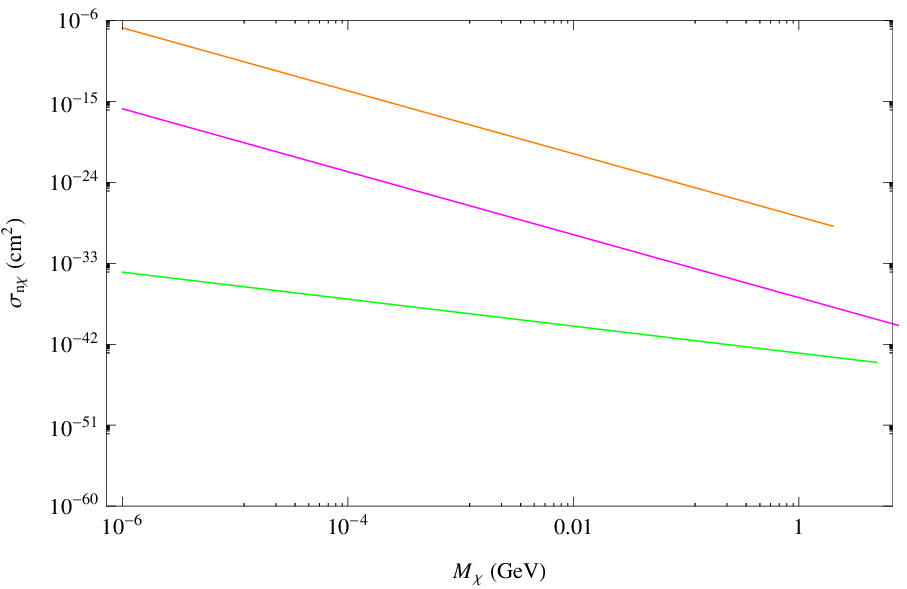}
	\caption{Same as in Fig.~\ref{fig:2} but for case II.}
	\label{fig:3}
\end{figure}

\begin{figure}[ht!]
	\centering
	\includegraphics[scale=1]{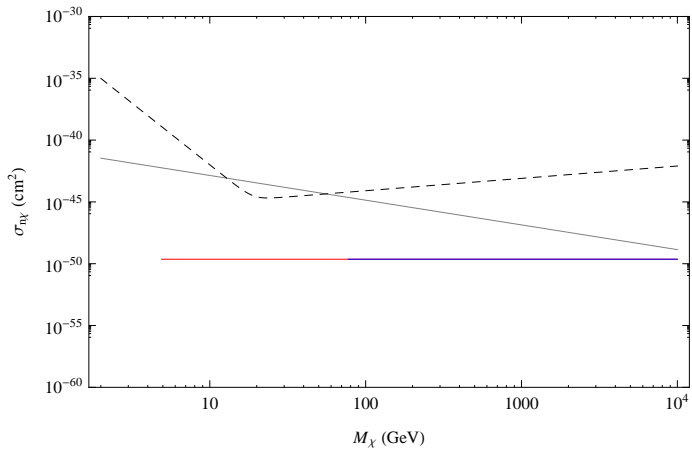}
	\caption{Same as in Fig.~\ref{fig:2} but for heavy DM particles. The gray curve corresponds to $\lambda_3$, the blue curve corresponds to $\lambda_2$, while the dashed curve corresponds to direct detection limits \cite{detection1,detection1b,detection2}.}
\label{fig:4}	
\end{figure}

\begin{figure}[ht!]
	\centering
	\includegraphics[scale=1]{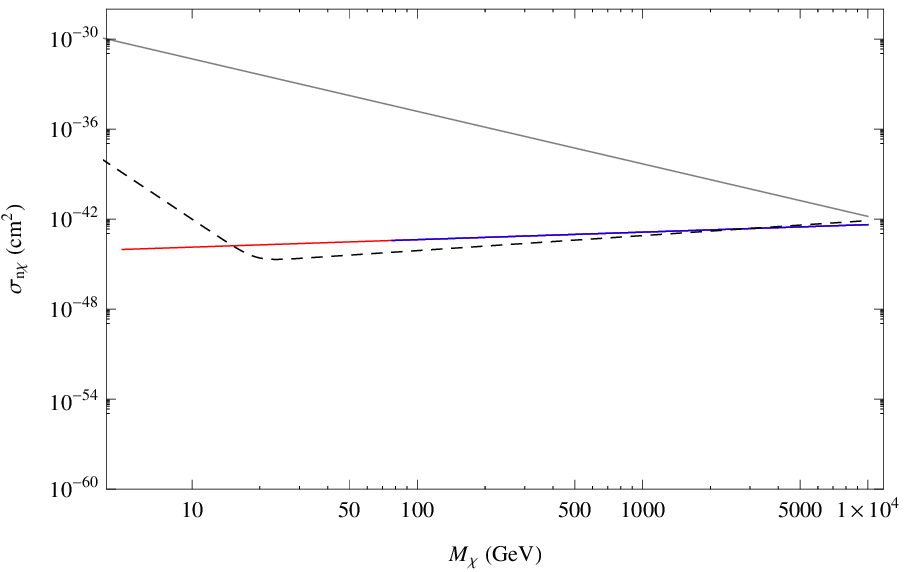}
	\caption{Same as in Fig.~\ref{fig:4} but for case II.}
	\label{fig:5}
\end{figure}

\begin{figure}[ht!]
	\centering
	\includegraphics[scale=1]{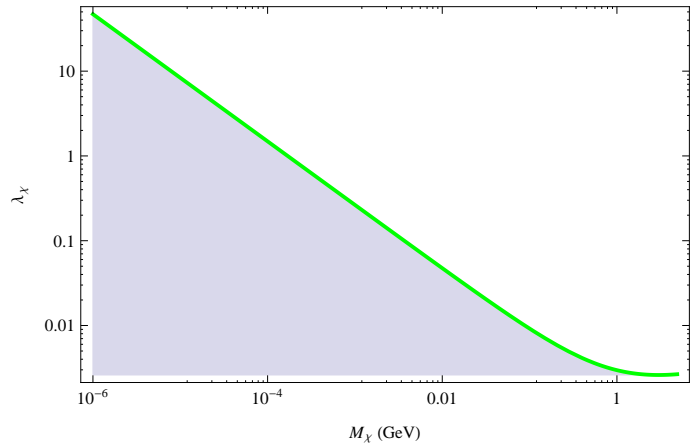}
	\caption{Same as in Fig.~\ref{fig:1} but the limit comes from the formation of the mini black hole inside the neutron star (for case II, light DM particles and negligible self-interactions).}
	\label{fig:6}
\end{figure}

First of all, given the conditions presented in the discussion above it is is easy to verify that:

a) Once the DM particles are thermalized the BEC will be formed.

b) For a DM particle mass in the range of interest $1keV < M_\chi < 10 TeV$, $r_{th}$ is lower than the radius of the star, so the DM particles are indeed trapped inside the neutron star. In addition, $r_c$ is lower than $r_{th}$ which implies that the DM particles are indeed concentrated inside a sphere with radius $r_c$ and not inside a sphere with radius $r_{th}$ as it is expected if the BEC is not formed.

c) For $\lambda_3$ the strongest condition comes from the uncertainty principle in the whole mass range, while for $\lambda_1, \lambda_2$ it depends on the mass of the DM
particle. When the DM particles are relatively light, $1keV < M_\chi < 4.9 GeV$ for $\lambda_1$ or $1keV < M_\chi < 77.3 GeV$ for $\lambda_2$, the strongest condition for the formation of the black hole comes from the uncertainty principle, namely $M_{\chi acc} > M_{cr}$, otherwise the condition becomes $M_{\chi acc} > M_2$ (the black hole is not evaporated due to Hawking radiation). Therefore we have considered 2 separate cases for light or heavy DM particles.

Fig.~\ref{fig:2} and ~\ref{fig:3} show the allowed parameter space on the $M_\chi-\sigma_{n\chi}$ plane for light DM particles and for the cases I and II
respectively, and for 3 values of the DM self-interaction coupling constant.
The green curve for $\lambda_1=0$, magenta curve for $\lambda_2=10^{-30}$, and orange curve for $\lambda_3=10^{-21}$).
For a given DM mass, the DM-nucleon cross section must lie below the solid curve to avoid the formation of the mini black hole inside the star. In this mass range there are no experimental limits.

Fig.~\ref{fig:4} and ~\ref{fig:5} show the allowed parameter space on the $M_\chi-\sigma_{n\chi}$ plane for heavy DM particles for the cases I and II respectively, and for 3 values of the DM self-interaction coupling constant.
The gray curve corresponds to $\lambda_3$, while the blue curve corresponds to $\lambda_2$. For a given DM mass, the DM-nucleon cross section must lie below the solid curve to avoid the formation of the mini black hole inside the star. For comparison we show in the same plot the limits from direct detection searches~\cite{detection1,detection1b,detection2} (dashed curve). Thus in the case II as well as when the self-interaction coupling constant is large in case I, neutron stars fail to provide us with bounds better than current limits from DM direct detection searches.

The allowed parameter space in the {$M_\chi-\lambda_\chi$} plane is shown in Fig.~\ref{fig:6} for case II, light DM particles in the mass range $1 keV < M_\chi < (a few) GeV$ and for negligible DM self-interactions. For a given DM mass, the coupling constant $\lambda_\chi$ must lie below the curve. As we can see the bound obtained here using neutron stars is comparable to the constraint coming from SM Higgs boson invisible decays \cite{invisible} shown in Fig.~\ref{fig:1}, unless the DM particles are extremely light, $M_\chi \leq 0.02 GeV$.

\section{Conclusions}

In the present article we have used for the first time neutron stars to constrain the parameter space of the scalar singlet and inert DM models. These new classes of DM candidates are the simplest extensions of the SM
of particle physics and very attractive. Since in the spectrum of these models there is a neutral scalar particle that is stable, massive and weakly coupled, it is a natural DM candidate. Indeed it has been shown that both the singlet scalar or the CP even Higgs boson in the inert model are excellent dark matter candidates.
The parameter space is simple and consists of two free parameters only, namely
the scalar mass $M_\chi$ and the dimensionless coupling constant $\lambda_\chi$. The latter determines the strength of the interaction of the DM particle to the SM Higgs boson. Given that neutron stars do exist we were able to
constrain the scalar DM parameter space by avoiding the formation of a mini black hole inside the star. Our findings indicate that i) for heavy DM particles neutron stars can provide us with bounds better that the current limits from direct detection searches only when the self-annihilations of DM particles are negligible and the DM self-interaction coupling constant is very small, while ii) for light DM particles the bounds obtained here are comparable to limits from Higgs invisible decays unless the DM particles are extremely light. Overall, although our study implies a significant reduction of the parameter space constrained by our analysis, resulting non-competitive with collider searches, it serves as a new and independent test.


\begin{acknowledgments}
We wish to thank the anonymous reviewer for his/her suggestions that helped us improve the quality of the manuscript. Our work was
supported from "Funda{\c c}{\~a}o para a Ci{\^e}ncia e Tecnologia".
\end{acknowledgments}


\end{document}